\documentclass[journal=jacsat,manuscript=article]{achemso}
\usepackage{longtable}
\usepackage{multirow}
\usepackage{amsmath}
\usepackage{subfigure}
\usepackage[usenames,dvipsnames]{color}
\SectionNumbersOn
\author{Marcos H. Gim\'{e}nez}
\affiliation{Departamento de F\'{\i}sica Aplicada, Universitat
Polit\`{e}cnica de Val\`{e}ncia, Camino de Vera s/n, 46022,
Valencia, Spain.}
\author{Isabel Salinas}
\affiliation{Departamento de F\'{\i}sica Aplicada, Universitat
Polit\`{e}cnica de Val\`{e}ncia, Camino de Vera s/n, 46022,
Valencia, Spain.}
\author{Juan C. Castro-Palacio}
\affiliation{Department of Earth Sciences and Engineering, Faculty
of Engineering, Imperial College London, London SW7 2AZ, United
Kingdom.}
\author{Jos\'{e} A. G\'{o}mez-Tejedor}
\affiliation{Centro de Biomateriales e Ingenier\'{\i}a Tisular,
Universitat Polit\`{e}cnica de Val\`{e}ncia, Camino de Vera s/n,
46022, Valencia, Spain.}
\author{Juan A. Monsoriu}
\email{jmonsori@fis.upv.es} \affiliation{Centro de Tecnolog\'{\i}as
F\'{\i}sicas, Universitat Polit\`{e}cnica de Val\`{e}ncia, Camino de
Vera s/n, 46022, Valencia, Spain.}
\date{\today}

\title{Visualizing acoustical beats with a smartphone}

\begin{document}

\begin{abstract}
In this work, a new Physics laboratory experiment on Acoustics beats
is presented. We have designed a simple experimental setup to study
superposition of sound waves of slightly different frequencies
(acoustic beat). The microphone of a smartphone is used to capture
the sound waves emitted by two equidistant speakers from the mobile
which are at the same time connected to two AC generators. The
smartphone is used as a measuring instrument. By means of a simple
and free Android$^{\rm TM}$ application, the sound level (in dB) as
a function of time is measured and exported to a .csv format file.
Applying common graphing analysis and a fitting procedure, the
frequency of the beat is obtained. The beat frequencies as obtained
from the smartphone data are compared with the difference of the
frequencies set at the AC generator. A very good agreement is
obtained being the percentage discrepancies within 1 \%.


\textit{Abstract in Spanish language}\\

El presente trabajo expone un nuevo experimento docente sobre el
Batido Ac\'{u}stico. En este sentido, hemos dise\~{n}ado un montaje
experimental para el estudio de la superposición de dos ondas de
sonido con frecuencias muy similares, lo que da origen al
fen\'{o}meno conocido como Batido Ac\'{u}stico. El micr\'{o}fono del
smartphone se utiliza para registrar las ondas de sonido emitidas
por dos altavoces, los que se ubican de manera equidistante con
respecto a este, y se conectan a generadores de corriente alterna.
El smartphone se utiliza aqu\'{i} como un instrumento de
medici\'{o}n. Los datos del nivel sonoro se capturan con ayuda de
una aplicaci\'{o}n Android$^{\rm TM}$, que se puede descargar de
Google Play de manera gratuita. Finalmente, haciendo uso de
t\'{e}cnicas de an\'{a}lisis gr\'{a}fico y de ajuste, se obtiene la
frecuecia del batido, la que a su vez se compara con las frecuencias
emitidas por los generadores. Los valores de discrepancia obtenidos
no superan el 1 \%.

Palabras claves: batido ac\'{u}stico, nivel de sonido, smartphone.

\end{abstract}

\section{Introduction}
Portable devices' sensors offer a wide range of possibilities for
the development of Physics teaching experiments in early years. For
instance, digital cameras can be used to follow physical phenomena
in real time since distances and times can be derived from the
recorded video\cite{monso:2005}. Wireless devices such as wiimote
have been also used in Physics teaching
experiments\cite{tomarken:2012}. The wiimote carries a three-axes
accelerometer which communicates with the console via
Bluetooth$^{\rm TM}$. More recently, smartphones have been
incorporated to this variety of portable
devices\cite{kuhn:2012,jccp:2013}. For instance, the acceleration
sensor of the smartphones has been used to study mechanical
oscillations, at both the qualitative\cite{kuhn:2012} and the
quantitative\cite{jccp:2013} levels. These works show very simple
experiments where the smartphone itself is the object under study.
The acceleration data are captured by the acceleration sensor of the
device and collected by the proper mobile app.

\noindent All smartphones are equipped with a microphone, which can
be used to record sounds with a sample rate of  44100 Hz, and in
some new devices up to 48000 Hz. This allows to analyze different
acoustic phenomena with the smartphone
microphone\cite{kuhn:2013a,hirth:2015,monteiro:2015}. The sound
frequency spectrum captured by the smartphone microphone can be
analized with a number of free applications, such as ``Audio
Spectrum Monitor"\cite{audiospectrum} and ``Spectrum
Analyzer"\cite{spectrumanalizer}. Also, the fundamental frequency of
a sound wave can be measured with very high precision, which allows
to study a frequency-modulated sound in Physics
laboratory\cite{tejedor:2015} and the Doppler Effect for sound waves
\cite{tejedor:2014}.

\noindent Using smarphone devices, several methods to measure the
velocity of sound have been proposed. For instance, by means of the
Doppler effect using ultrasonic frequency and two smartphones the
speed of sound can be determined with an accuracy of about $5 \%$
\cite{klein:2014}. Based on the distance between the two smartphones
and the recording of the delay between the sound waves traveling
between them, the actual speed of sound can be
obtained\cite{parolin:2013}. Using economic instruments and a couple
of smartphones, it is possible to see nodes and antinodes of
standing acoustic waves in a column of vibrating air and to measure
the speed of sound\cite{parolin:2015}. By the study of destructive
interference in a pipe it is also possible to adequately and easily
measure the speed of sound\cite{yavuz:2016}. A soundmeter
application can be used to measure the resonance in a beaker when
waves with different wavelengths are emitted by the smartphone
speaker. This application can also be used to measure and analyze
Doppler effect, interferences, frequencies spectra, wavelengths,
etc. or to study other phenomena in combination with some other
fundamental physics laboratory equipment such as Kundt or Quincke
tubes\cite{gonzalez:2015}. On the other hand, measurements with the
smartphone microphone can be used to analyze physical process not
directly related with acoustic. The sounds made by the impacts of a
ball can be recorded with the microphone. The impacts resulting in
surprisingly sharp peaks that can be seen as time markers. The
collected data allows the determination of gravitational
acceleration \cite{schwarz:2013}.

\noindent In order to measure the acoustic beat, two mobile phones
can be placed at a short distance from each other and then play
previously recorded tones with a constant frequency with the MP3
function. The signal can be captured using a microphone and by the
line-in of a sound card in a computer, and then, the recorded signal
can be analyzed with suitable audio software\cite{kuhn:2013b}. In a
similar way, three smartphones can be used to analyze the acoustic
beat: two of them produce the sine tones with slightly different
frequencies and the third device detects and analyzes the
overlapping oscillation\cite{kuhn:2014}. In this kind of
experiments, oscillogrammes are recorded and the acoustic beats are
derived from the varying envelope amplitude.

\noindent In the present work, a more intuitive procedure is
presented in order to characterize acoustic beats with a smartphone.
When two sound waves of very close frequencies are superimposed, a
``vibrating" tone is perceived. This is the basic principle behind
the tuning of musical instruments. Instead of using oscillogrammes,
we propose to capture the perceived vibrating tone by using the
smartphone as a sonometer. The sound waves are generated by two
independent speakers connected to AC generators, although two other
smartphones may be also used to generate the sine tones. By means of
the free App. ``Physics Toobox Sound Meter" \cite{phystool}, the
students are able to measure the sound intensity (in dB) of the
acoustic beats as a function of time.

\noindent The resulting sound intensity variations are directly
displayed on the mobile screen and the frequency beat can be
quantitatively obtained. Moreover, the recorded sound levels can be
exported to a PC for a more quantitative analysis, i.e. by email,
cable or bluetooth connection. In this way, the varying intensity of
the vibrating tone is derived from the sound level measurements and
fitted to a harmonic function in order to accurately obtain the
corresponding frequency beat. The results are compared with the
frequency difference of the superimposing AC signals, and a very
good agreement is obtained.

\section{Basic theory}
Let  $x_{\rm 1}(t)$ and $x_{\rm 2}(t)$ be two harmonic oscillations
of equal amplitude $A$, very close frequencies $f$ y $f+\Delta f$,
and initial phases $\varphi_1$ and $\varphi_2$,
\begin{equation}
x_1(t)=Asin[2\pi f t+\varphi_1],
 \label{eq1}
\end{equation}
\begin{equation}
x_2(t)=Asin[2\pi (f+\Delta f)t+\varphi_2].
 \label{eq2}
\end{equation}
After some basic mathematical manipulations, the superposition of
both oscillations gives rise to,
\begin{equation}
x=x_{\rm 1}(t)+x_{\rm 2}(t) = A'sin[2\pi (f+\frac{\Delta
f}{2})t+\frac{\varphi_2 + \varphi_1}{2}].
 \label{eq3}
\end{equation}

 \noindent
The frequency of the resulting oscillation is the average value of
the superimposing oscillations. The resulting amplitude is,
\begin{equation}
A'=2Acos(2\pi\frac{\Delta f}{2}t + \frac{\varphi_2 - \varphi_1}{2}).
 \label{eq4}
\end{equation}

\begin{figure}[H]
\centering
\includegraphics[scale=0.45]{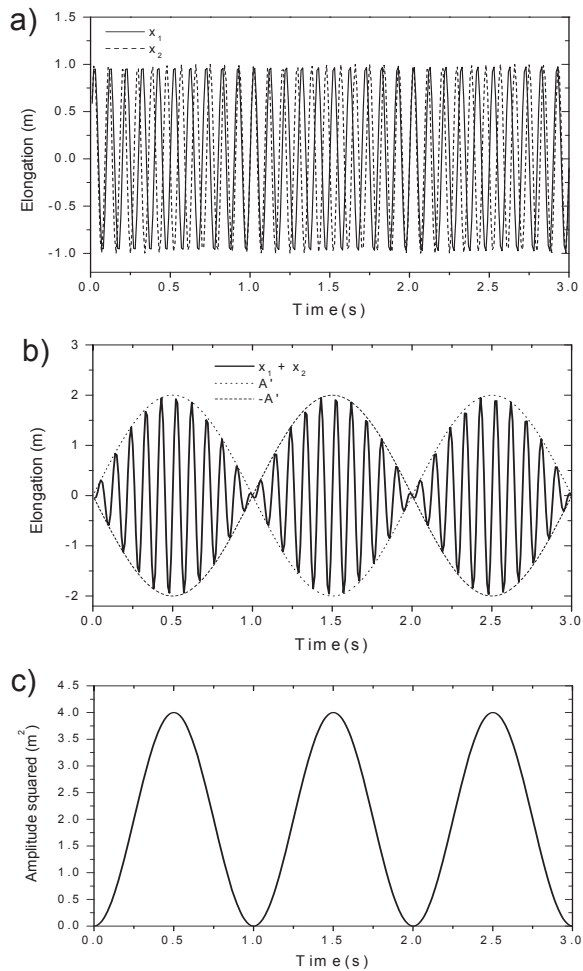}
\caption{Superposition of two oscillations $x_{\rm 1}(t)$ and
$x_{\rm 2}(t)$ of the same amplitude ($A=1$ m) and close frequencies
($f=10$ Hz ; $f+\Delta f=10+1=11$ Hz). In panel a), the single
oscillations are shown. In panel b), the superposition of the
oscillations and envelope curve and in panel c), the amplitude
squared.} \label{fig:fig1}
\end{figure}

 \noindent
The intensity of the wave resulting from the interference of the
initial oscillations is proportional to the amplitude squared. Let
us denote the proportionality factor as $\lambda$,
\begin{equation}
I=\lambda A'^2 = 4 \lambda A^2 cos^2(2\pi\frac{\Delta
f}{2}t+\frac{\Delta \varphi}{2}) = 2\lambda A^2+2\lambda
A^2cos[2\pi\Delta ft+\Delta \varphi]. \label{eq5}
\end{equation}
where $\Delta \varphi = \varphi_2 - \varphi_1$. The above equation
can be rewritten as,
\begin{equation}
I=\frac{I_{max}}{2} [ 1 + cos[2\pi f_bt + \Delta \varphi ].
 \label{eq55}
\end{equation}
where $f_b= \Delta f$ is the frequency beat and $I_{max}$ is the
maximum sound intensity. Therefore, the resulting frequency of
oscillation is the difference of the interfering oscillations.
Figure \ref{fig:fig1} shows the example of $x_{\rm 1}$(t) and
$x_{\rm 2}$(t) which are oscillations of the same amplitude, $A=1$
m, close frequencies $f=10$ Hz and $f+\Delta f=10+1=11$ Hz, and
initial phases $\varphi_1=\varphi_2=0$ rad. The frequency of the
amplitude squared $A'$ (and so of the intensity $I=\lambda A'^2$) is
$\Delta f=1$ Hz.

 \noindent
All the theory explained above is applicable to sound waves such as
those generated by speakers placed at equal distances from the
microphone of the smartphone. The speakers are feeded with slightly
different signals of same effective voltage from two independent AC
generators.

\section{Experimental results} 
\begin{figure}[H]
\centering
\includegraphics[scale=0.45]{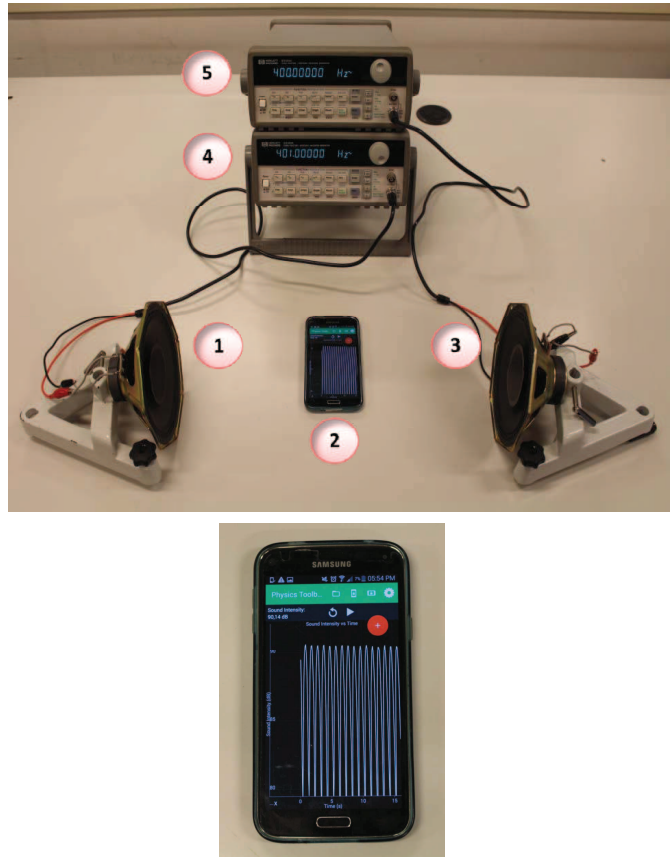}
\caption{In the upper panel, a photograph of the experimental setup
is shown. In the figure, the speakers (1 and 3), the smartphone (2)
and the AC generators (4 and 5) are shown. In the lower panel,
variations of the sound level of the acoustic beat as shown on the
smartphone´s screen by the application ``Physics Toolbox Sound
Meter".} \label{fig:fig2}
\end{figure}

The experimental setup used to carry
out the experiments is shown in Figure \ref{fig:fig2}a. It consists
of two AC generators (model 33120 A of Hewlett Packard), two
identical speakers (model AD70800/M4 from Philips) facing each
other, and the necessary cables to get all appliances connected.
Finally, the smartphone is placed in the mid-way between both
speakers. Two other smartphones with an App. for generating a sine
tone, could be also used. The Android application ``Physics Toolbox
Sound Meter", capable of measuring the sound level of the waves
coming from the speakers was previously installed on the smartphone.

 \noindent
First, the same effective voltage is set at both AC generators. The
speakers were feeded with signals of similar frequency and within
the human audible range. We have used the frequencies 400 Hz and 401
Hz in the example shown in Figure \ref{fig:fig2} (lower panel).
After checking that the beat can be heard, the mobile application is
turned on. The beat oscillations are then observed on the mobile
screen (Figure \ref{fig:fig2}b). It can be verified that, even when
there is a small level of background noise, and the sampling
frequency can not resolve the minimum values of the signal, the
periodicity of the oscillations are still observed.
\begin{figure}[H]
\centering
\includegraphics[scale=0.45]{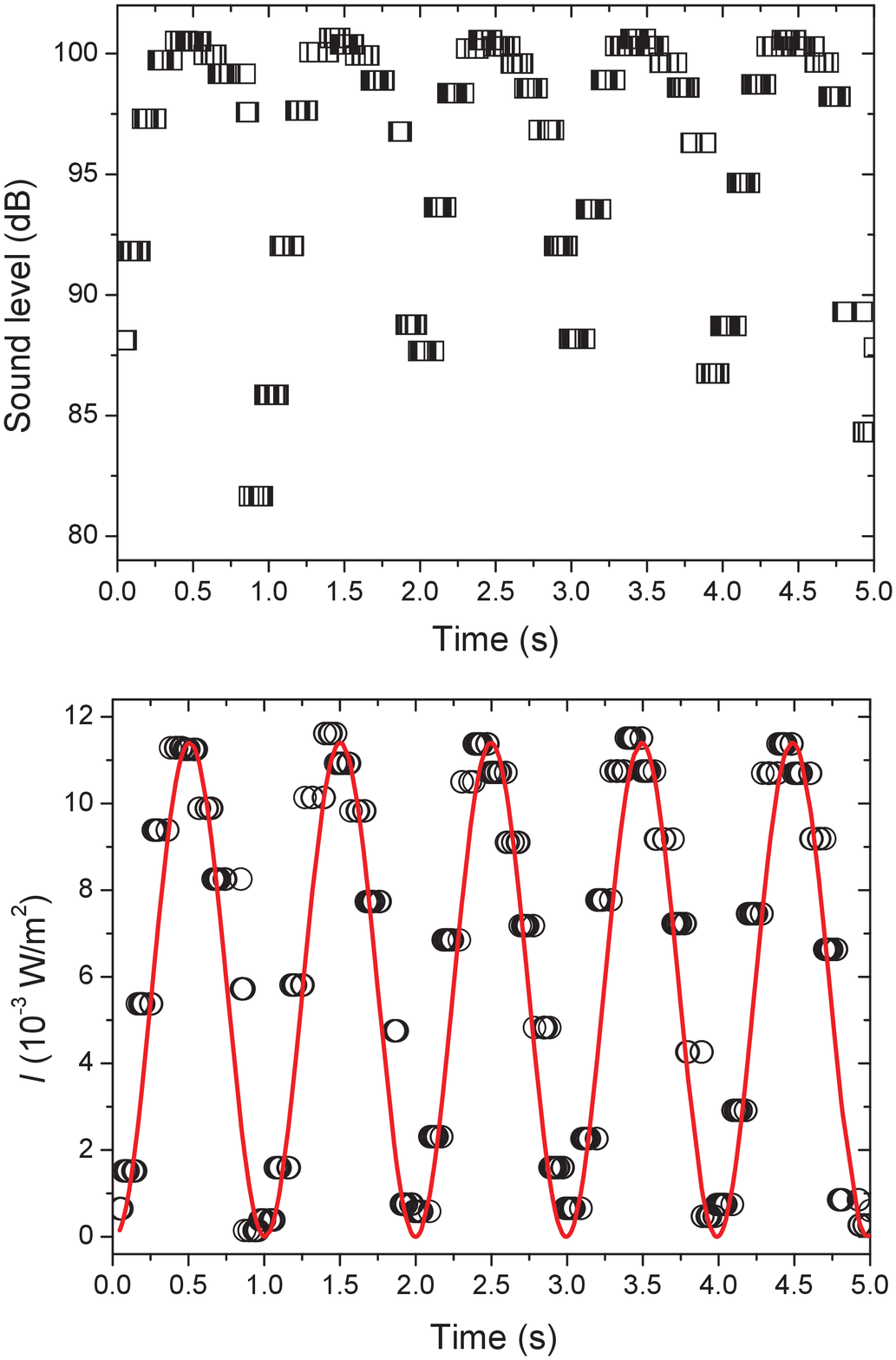}
\caption{In panel a), the time series of the sound level in dB
($\beta$) (open squares) is shown and in panel (b), the time series
of the sound intensity ($I(t)$ in Eq.\ref{eq5}) (open circles),
along with the fitted function (red solid line). Results correspond
to a frequency difference of 1 Hz in the AC generators.}
\label{fig:fig3}
\end{figure}
\noindent After recording the sound level for several seconds, the
registered data, previously exported to a .csv file, can be sent to
a PC for further analysis. For this purpose, different ways can be
used, namely, cable connection, bluetouth or email. In order to
derive the beat frequency, the first step is to convert the
registered sound level $\beta$ in dB to the sound intensity $I$ in
W/m$^{\rm 2}$ using the following expresion,\cite{tipler:2005}
\begin{equation}
I=I_010^{\beta/10}.
 \label{eq6}
\end{equation}
where $I_0=10^{-12}$ W/m$^{\rm 2}$ is the standard value of the
intensity threshold of the audible range in humans. Later, an
interval of 5 s is chosen from the central part of the time series
recorded by the smartphone. This segment of data for $I(t)$ is
fitted using a Least-squares algorithm to the Eq. \ref{eq55}. The
only relevant quantity to this work is the beat frequency, $f_b$,
although the other two parameters ($I_{max}$ and $\Delta \varphi$)
can be also obtained from the fitting procedure.

 \noindent
The use of sine or cosine functions in the fitting does not make any
difference since it affects only the initial phase and not the
frequency of the acoustic beat which is our objective function.
Based on the values of frequency obtained from the fitting, the
frequency of the beat $f_{\rm b}$ can be determined and compared
with $\Delta f$ which is the difference of the frequencies from the
AC generators.

 \noindent
Figure \ref{fig:fig3} shows the results for a frequency difference
in the AC generator as 1 Hz. First, the central interval of the time
series of the sound level (in dB), registered with the smartphone is
represented in Figure \ref{fig:fig3}a of $R^2=0.90$. The resulting
beat frequency is $f_b=(1.008 \pm 0.002)$ Hz which corresponds to a
discrepancy with respect to the frequency difference in the AC
generator of 0.8 \%.

\begin{figure}[H]
\centering
\includegraphics[scale=0.45]{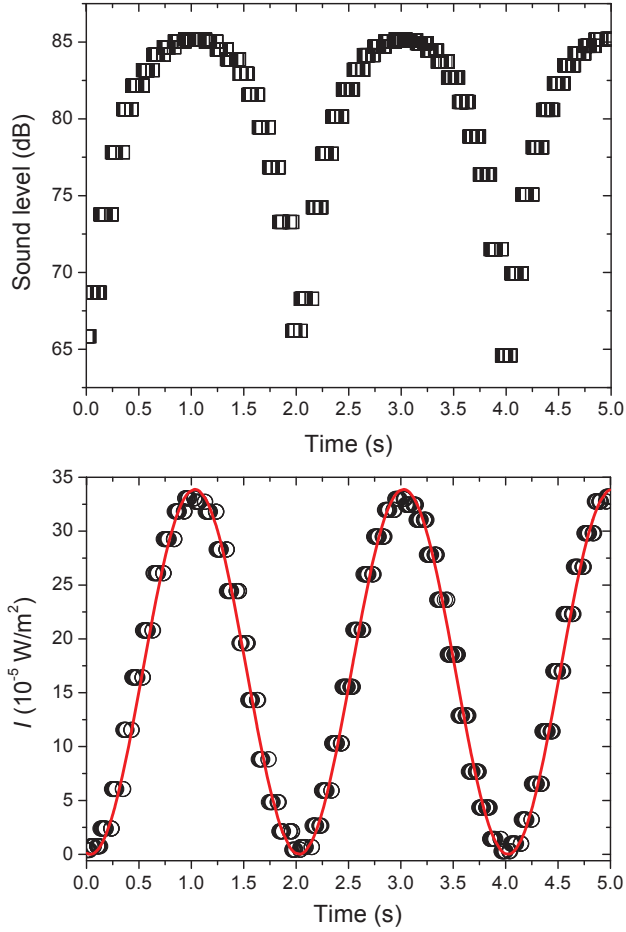}
\caption{In panel a), the time series of the sound level in dB (open
squares) is shown and in panel (b), the time series of the sound
intensity ($I(t)$ in Eq.\ref{eq5}) (open circles), along with the
fitted function (red solid line). Results correspond to a frequency
difference of 0.5 Hz in the AC generators.} \label{fig:fig4}
\end{figure}

 \noindent
In order to provide further verification of the experimental
procedure, several combinations of close frequencies
($f_b\approx\Delta f$) are used. For example, figure \ref{fig:fig4}
shows the results for frequency differences in the AC generators of
$\Delta f$ = 0.5 Hz, where an experimental frequency beat of
$f_b=(0.5020 \pm 0.0004)$ Hz with a regression coefficient
$R^2=0.99$. The quality of the fit can be seen in the value of $R^2$
which is close to 1. In this case, the discrepancy with respect to
the expected value is 0.4 \%.

 \noindent
 We repeated the proposed experimental procedure to characterize the
 acoustic beats with frequency differences in the AC generator
 between
 0.5 and 1.5 Hz and main frequencies between 400 and 700 Hz. In all
 cases, the discrepancies were lower than 1 \%. Therefore, the
 precision of the frequency measurements was reasonably enough to
capture the acoustic beat phenomenon. This is not itself the goal of
this work but to show the students a Physics teaching experiment
based on a smartphone, a very familiar device to them.

\section{Conclusions}
A new Physics teaching experiment for first years university
students has been presented in this work. A smartphone with an
Android$^{\rm TM}$ application has been used as sonometer to measure
the sound intensity of the beats formed by the superposition of
sound waves generated at speakers connected to AC generators. The
interfering waves had the same amplitude and very close frequencies.
The analysis of the time series generated from the measurements with
the smartphone are further analyzed by the students in order to
determine the frequency of the beats. The beat frequency obtained
from the smartphone data reproduces the value calculated from the AC
generators frequencies within 1 $\%$. The use of smartphones in
Physics teaching experiments is a very motivating experience for the
students. This has come up in our Physics courses where the students
have experimented with different types of
phenomena.\cite{jccp:2013,tejedor:2015,tejedor:2014}

\section*{Acknowledgments}
The authors would like to thank the Institute of Education Sciences
of the Polytechnic University of Valencia for the support given to
the research groups on teaching innovation: MoMA and MACAFI, and for
supporting the project PIME/2015/B18 which gave rise to this work.


\end{document}